\documentclass[draftclsnofoot,12pt,onecolumn]{IEEEtran}
\usepackage{amsmath,amsfonts,amsthm}
\usepackage{amssymb}
\usepackage{cite}

\usepackage[dvips, final]{graphicx}
\graphicspath{{./eps/}} \DeclareGraphicsExtensions{.eps}
\newcommand{\figwidth}{.8\linewidth}

\normalsize
\usepackage{color}

\newtheorem{theorem}{Theorem}
\newtheorem{lemma}{Lemma}
\newtheorem{result}{Result}
\def\lemref#1{Lemma~\ref{#1}}
\def\reref#1{Result~\ref{#1}}
\def\figref#1{Figure~\ref{#1}}
\def\thref#1{Theorem~\ref{#1}}
\newcounter{remark}
\def\remark{\addtocounter{remark}1\noindent\emph{Remark \arabic{remark}:} }
\newcounter{testcase}
\def\testcase{\addtocounter{testcase}1\smallskip%
\noindent\emph{Test Case \arabic{testcase}:} }
\def\rank{\text{rank}}
\def\trace{\text{Tr}}
\def\Re{\text{Re}}

\def\bA{{\mathbf A}}
\def\bB{{\mathbf B}}
\def\bC{{\mathbf C}}
\def\bI{{\mathbf I}}
\def\bP{{\mathbf P}}
\def\bU{{\mathbf U}}
\def\bX{{\mathbf X}}
\def\bY{{\mathbf Y}}
\def\bZ{{\mathbf Z}}
\def\bh{{\mathbf h}}
\def\br{{\mathbf r}}
\def\bs{{\mathbf s}}
\def\bu{{\mathbf u}}
\def\bv{{\mathbf v}}
\def\bx{{\mathbf x}}
\def\by{{\mathbf y}}
\def\bz{{\mathbf z}}
\def\bzero{{\mathbf 0}}

\def\cD{{\mathcal D}}

\def\tLambda{{\Tilde\Lambda}}
\def\bbs{{\Bar{\mathbf s}}}
\def\tbs{{\Tilde{\mathbf s}}}
\def\tbu{{\Tilde{\mathbf u}}}
\def\tbx{{\Tilde{\mathbf x}}}
\def\tby{{\Tilde{\mathbf y}}}
\def\ttau{{\Tilde\tau}}
\def\hbs{{\Hat{\mathbf s}}}
\def\hs{{\Hat s}}
\def\hI{{\Hat I}}

\begin{document}
\setcounter{page}{1}
\title{Multiple Access for Small Packets Based on Precoding and Sparsity-Aware
Detection\footnote{Part of this work will be presented at the IEEE/CIC
International Conference on Communications in China, 2014\cite{re8}.}}

\author{
Ronggui Xie$^1$, Huarui Yin$^1$, Xiaohui Chen$^1$, and Zhengdao Wang$^2$\\
\small $^1$Department of Electronic Engineering and Information Science,
University of Science and Technology of China\\ \small $^2$Department of
Electrical and Computer Engineering, Iowa State University}

\maketitle

\begin{abstract}
Modern mobile terminals often produce a large number of small data packets.
For these packets, it is inefficient to follow the conventional medium access
control protocols because of poor utilization of service resources. We propose
a novel multiple access scheme that employs block-spreading based precoding at
the transmitters and sparsity-aware detection schemes at the base station. The
proposed scheme is well suited for the emerging massive multiple-input
multiple-output (MIMO) systems, as well as conventional cellular systems with
a small number of base-station antennas. The transmitters employ precoding in
time domain to enable the simultaneous transmissions of many users, which
could be even more than the number of receive antennas at the base station.
The system is modeled as a linear system of equations with block-sparse
unknowns. We first adopt the block orthogonal matching pursuit (BOMP)
algorithm to recover the transmitted signals. We then develop an improved
algorithm, named interference cancellation BOMP (ICBOMP), which takes
advantage of error correction and detection coding to perform perfect
interference cancellation during each iteration of BOMP algorithm. Conditions
for guaranteed data recovery are identified. The simulation results
demonstrate that the proposed scheme can accommodate more simultaneous
transmissions than conventional schemes in typical small-packet transmission
scenarios.
\end{abstract}

\begin{IEEEkeywords}
small packet, block-sparsity, compressive sensing, massive MIMO, BOMP,
precoding, interference cancellation
\end{IEEEkeywords}

\IEEEpeerreviewmaketitle

\section{Introduction} As intelligent terminals such as smart phones and
tablets get more popular, they produce an increasing number of data packets of
short lengths, to be delivered over a cellular network. Modern mobile
applications that produce such small packets include instant messaging, social
networking, and other services \cite{re2},\cite{re3}. Although the lengths of
messages are relatively short, small packet services put great burden on the
communication network. Two kinds of messages contribute to the traffic of
small packets: one is the small packets of conversation produced by active
users that occupy only a small percentage of the total online users
\cite{re3}; the other is the signaling overheads needed to transmit these
conversation packets \cite{re4}.

In current wireless communication systems, a user follows the medium access
control (MAC) protocols to obtain the service resources. Two flavors of MAC
protocols are used in general: i) resource reservation based, and ii)
collision resolution based. In the first kind, resources are preallocated to
the users in a noncompetitive fashion. For small and random packets, the
reservation-based approach is inefficient in resource utilization due to
irregularity of the packets. In the collision-resolution based approaches, the
terminals are allowed to access the resources in arbitrary order and when
collision occurs, certain resolution mechanism is then employed. The
collision-resolution based MAC can suffer from too many retransmissions due to
frequent collisions.

In this paper, we propose a novel uplink small packet transmission scheme
based on precoding at the transmitters and sparsity-aware detection at the
receiver. The main motivation is to allow for a large number of users to
transmit simultaneously, although each user may be transmitting only a small
amount of data. Besides frame-level synchronization, no competition for
resources or other coordinations are required. This saves the signaling
overhead for collision resolution, and improves the resource utilization
efficiency.

The contributions of our work are as follows:
\begin{enumerate}

\item \emph{Block precoding and block-sparse system modeling:} We apply block
precoding at each transmitter in time domain, and by considering the user
activities, develop a block-sparse system model that takes full advantages of
the structure of the signals to recover and is suitable for
compressive-sensing based detection algorithms.

\item \emph{Sparsity-aware detection algorithm:} We develop a interference
cancellation (IC) based block orthogonal matching pursuit (ICBOMP) algorithm.
The algorithm improves upon the traditional BOMP algorithm by taking advantage
of availability of error correction and detection, which is common in digital
communications. By ICBOMP algorithm, not only do we achieve much better signal
recovering accuracy but we also benefit in terms of less computational
complexity. The price is slightly decreased rate due to coding.

\item \emph{Signal recovery conditions:} We derive conditions for guaranteed
signal recovery. The condition we require on the BOMP algorithm is milder than
that in the related work in \cite{re24}. For ICBOMP algorithm, we give the
conditions for perfect IC in each iteration. Furthermore, we characterize the
data recovery condition from information theoretic point of view.

\end{enumerate}

Thanks to the precoding operation and our sparsity-aware detection algorithms,
our scheme enable the system to support more active users to be simultaneously
served. The number of active users can be even more than the number of
antennas at base station (BS). This is of great practical significance for
networks offering small packet services to a large number of users.

Our proposed scheme is especially suitable for the so called massive
multiple-input multiple-output (MIMO) systems \cite{re9}-\cite{re12}. In
massive MIMO systems, the number of antennas at the BS can be more than the
number of active single-antenna users that are simultaneously served. When the
number of antennas at BS is large, the different propagation links from the
users to the BS tend to be orthogonal, and the large amount of spatial degrees
of freedom are useful for mitigating the effect of fast fading
\cite{re10},\cite{re11}. Overall, massive MIMO technique provides higher data
rate, better spectral and energy efficiencies \cite{re12}.

Applications of compressive sensing (CS) to random MAC channels have been
considered in \cite{re35}-\cite{re39}. In \cite{re35}, CS based decoding
scheme at the BS has been used for the multiuser detection task in
asynchronous random access channels. A technique based on CS for meter reading
in smart grid is proposed in \cite{re37}, and its consideration is limited to
single-antenna systems. Besides, a novel neighbor discovery method in wireless
networks with Reed-Muller Codes has been proposed in \cite{re39}, where CS
technique is also adopted. All the referred works depend on the idea that the
MAC channel is sparse, and all their works are classified to initial category
of CS, where no structure property have been taken into account. This is one
of the main distinctions that differentiate our work from the referred ones.

The rest of the paper is organized as follows. In Section \ref{SystemModel},
the system model of block sparsity are given. In Section \ref{BOMP}, we
introduce the BOMP algorithm and its improved version ICBOMP algorithm to
recover the transmitted signals. Guarantees for data recovery are presented in
Section \ref{Guarantees}. Section \ref{simulations} will present the numerical
experiments that prove the effectiveness of our scheme. Afterwards, to invest
the scheme with practical significance, some issues are discussed in Section
\ref{Discussion}. Finally, the conclusion will be presented in Section
\ref{Conclusion}. To better organize the contents, we will relegate some of
the proofs to the appendix.

Notation: Vectors and matrices are denoted by boldface lowercase and uppercase
letters, respectively. The 2-norm of a vector $\bv$ is denoted as ${\| \bv
\|_2}$, and the 0-norm is given as ${\| \bv \|_0}$. The inner product of two
vectors $\bv_1$ and $\bv_2$ is denoted as $( {{\bv_1},{\bv_2}})$. The identity
matrix of $d \times d$ dimension is denoted as $\bI_d$. For a given matrix
$\bU$, its conjugate transpose, transpose, pseudo inverse, trace and rank are
respectively denoted as $\bU^H$, $\bU^T$, ${\bU^\dag }$, $\trace\{\bU \}$,
$\rank\{ \bU \}$, and the spectral norm of $\bU$ is given by $\| \bU \|$.
Operation $vec(\bU)$ denotes vectorizing $\bU$ by column stacking. For a
subset $I\subset [N]:=\{1,2, \cdots ,N\}$ and matrix $\bU: = [
{{\bU_1},{\bU_2}, \cdots ,{\bU_N}}]$ consisting of $N$ sub-matrices (blocks),
where each sub-matrix has equal dimensionality, $\bU_I$ denotes a sub-matrix
of $\bU$ with block indices in $I$; for a vector $\bv: = {[ {\bv_1^T,\bv_2^T,
\cdots ,\bv_N^T} ]^T}$, $\bv_I$ is similarly defined. For a set $A$, $|A|$
denotes its cardinality. For two sets $A$ and $B$, $A\backslash B:=A\cap B^c$
denotes the set difference. For a real number $r$, $|r|$ and $\Re(r)$, and
$\lfloor r \rfloor $ denote its absolute value, real part, and floor,
respectively. Operation $\otimes$ denotes the Kronecker product of two
matrices.

\section{System Model}\label{SystemModel}

Consider an uplink system with $N$ mobile users, each with a single antenna,
and a BS with $M$ antennas. When a terminal is admitted to the network, it
becomes an online user. We assume that there are $N_a$ active users, out of
the total $N$ online users, that have data to transmit. It is not required
$N_a$ be known a priori or ${N_a} < M$; actually, in practical systems $N_a$
is usually unknown and it is possible that $N_a \gg M$.

We make the following further assumptions on the system considered.
\begin{enumerate}
\item The channels are block-fading: it remains constant for a certain
duration and then changes independently.

\item The transmissions are in blocks and the users are synchronized at the
block level. We assume that each frame of transmission consists of $T$
symbols, which all fall within one channel coherent interval.

\item The users each have single antenna. There are multiple antennas at the
BS.

\item The antennas at the BS, as well as the antennas among users, are
uncorrelated and uncoupled.

\item The BS always has perfect channel state information (CSI) of online
users.
\end{enumerate}

Let $\bs_n \in {\mathbb{C}^{d \times 1}}$ denotes the symbols to be
transmitted by user $n$, with $d < T$. User $n$ applies a precoding to $\bs_n$
to yield
\begin{equation}
{\bx_n} = {\bP_n}{\bs_n}
\end{equation}
where $\bP_n$ is a complex precoding matrix of size $T\times d$. The entries
of $\bx_n$ are transmitted in $T$ successive time slots. The received signals
at all antennas within one frame can be written as
\begin{equation}
\bY = \sqrt {{\rho_0}} \sum_{n = 1}^N {{\bh_n}\bx_n^T} +\bZ
=\sqrt {{\rho_0}} \sum_{n = 1}^N {{\bh_n}\bs_n^T{\bP_n^T}} +\bZ
\end{equation}
where ${\rho_0}$ is the signal to noise ratio (SNR) of the uplink, $\bY$ is
noisy measurement of size $M\times T$, $\bZ \in {\mathbb{C}^{M \times T}}$
represents the additive noise, with i.i.d.\ circularly symmetric complex
Gaussian distributed random entries of zero mean and unit variance, and $\bh_n
\in {\mathbb{C}^{M \times 1}}$ represents the channel coefficients from the
user $n$ to the base station, without loss of generally, let
${h_{mn}}\sim{\mathcal{CN}}({0,1})$, $m = 1,2, \cdots ,M$. Using the linear
algebra identity $vec({\bA\bB\bC}) = ({{\bC^T} \otimes \bA})vec(\bB)$, we can
rewrite the received signal as
\begin{equation} \label{model0}
vec(\bY) =\sqrt {{\rho_0}} \sum_{n = 1}^N {(\bP_n \otimes {\bh_n}){\bs_n}}
 +vec(\bZ)
\end{equation}

Define $\by: = vec(\bY)$, ${\bB_n}: = ({\bP_n \otimes {\bh_n}})/\sqrt M $ and
$\bB: = [ {{\bB_1},{\bB_2}, \cdots ,{\bB_N}} ]$, $\bs: = {[ {\bs_1^T,\bs_2^T,
\cdots ,\bs_N^T} ]^T}$. Then we can write the model in \eqref{model0} as
\begin{equation} \label{model1}
{\by} =\sqrt {{\rho_0}M} {\bB}{\bs}+ \bz
\end{equation}

In this formulation, we have assumed that all the users have messages of equal
length $d$. This may not be the case in practice. We view $d$ as the maximum
length of the messages of all users within a frame. For the users whose
message length is less than $d$, we assume their messages have been
zero-padded to $d$ before precoding. Also, for those users that are not
active, we assume their transmitted symbols are all zeros.

Model \eqref{model1} indicates that the signals to recover present the
structure of block-sparsity where transmitted signals are only located in a
small fraction of blocks and all other blocks are zeros. The length of each
block is $d$. We collect all the indices of blocks corresponding to active
users to form a set $I$, with $| I | = N_a \le K$, which means the unknown
number of nonzero blocks (active users) is at most $K$. Our consideration is
limited to case $MT < Nd$, where \eqref{model1} represents an under-determined
system. For case where $MT > Nd$, the receiver design is easier and the
proposed method is also applicable. When precoding matrix ${\bP_n}$ is
reasonably designed, matrix ${\bB}$ can meet the requirement for sensing
matrix in CS, and this kind of ${\bP_n}$ is of wide range, for instance,
Gaussian or Bernoulli matrix. Therefore, model \eqref{model1} can be viewed as
block sparsity model in CS \cite{re23,re24}. In CS, ${\bB}$ is referred as
dictionary.

\remark The following are several remarks on our precoding scheme:
\begin{enumerate}
\def\labelenumi{\arabic{remark}\alph{enumi})}

\item The precoding scheme is proposed because in reality, $T$ is usually
several times longer than the lengths of small packets. Also, the precoding
scheme contributes to solving signal recovery problem in the situation where
$N_a>M$.

\item Each user knows its own precoding matrix and the BS knows all precoding
matrices of all users.

\item A basic requirement on the precoding matrix is that it should be full
column rank, which is a requirement for data recovery. Additionally, in order
to balance the power of every symbol of the messages before and after being
precoded, each column of ${\bP_n}$ should be normalized to unit energy.

\item Our precoding scheme is different from spreading schemes in \cite{re37},
\cite{re38}, where direct sequence spread spectrum (DSSS) is utilized for CS
formulation.

\end{enumerate}

\section{Algorithms For Data Recovery}\label{BOMP}

The past few years have also witnessed the research interest in CS
\cite{re13,re14,re15,re16}. Initial works in CS treat sparse weighting
coefficients as just randomly located among possible positions in a vector.
When structure of the sparse signal is exploited, for example block sparsity,
it is possible to obtain better signal reconstruction performance and reduce
the number of required measurements \cite{re20,re23,re24,re25}.

We develop detection algorithms to be used at the BS in this section. We first
apply known algorithm BOMP to our problem, and then further improve it by
incorporating IC based on error correction and detection.

\subsection{BOMP Algorithm} The main idea of BOMP algorithm is that, for each
iteration, it chooses a block which has the maximum correlation with the
residual signal, and after that, it will use the selected blocks to
approximate the original signals by solving a least squares problem
\cite{re24}. For later convenience, we present the details of BOMP algorithm
as follows:

\begin{enumerate}
\item \emph{Input:} Matrix ${\bB}$, signal vector $\by$, ${\rho_0}$, $M$, $T$
and $d$.

\item \emph{Parameter setting:} Maximum number of iterations $K$. Usually, $K
\le \lfloor \frac{MT}d \rfloor $. With $K$ iterations, at most $K$ active
users can be identified.

\item \emph{Initialization:} Index set ${{\Lambda}_t} = \O $, basis function
set ${\mathbf{\Theta}_0} = \O $, residual signal ${\br_0} = \by$, the number
of iterations $t = 1$.

\item \emph{Main iteration:} While $t \le K$, do the following
  \begin{enumerate}
  \def\labelenumii{\arabic{enumi}\alph{enumii})}
  \item Calculate the of correlation coefficients given by the
  residual signal with each column of ${\bB}$, denoting as
  ${{\bB}^H}{\br_{t-1}}$.

  \item Find the index ${\lambda_t} \in \{ {1,2, \ldots
  ,N} \}$ of the block and the block unit $\bB_{{\lambda_t}}$ of
  ${\bB}$, satisfying $\{ {{\lambda_t},\bB_{\lambda_t}}
  \} = \arg \max_{j \in \{ {1, \cdots ,N} \}}
  {\| {{\bB_j^H}{\br_{t-1}}} \|_2}$.

  \item Augment the index set and basis function set, and set the
  ${\lambda_t}$-th block of ${\bB}$ a zero sub-matrix \[{{\Lambda}_t}
  = {{\Lambda}_{t-1}} \cup \{ {\lambda_t}\}\] \[\mathbf{\Theta}_t
  = [ {{\mathbf{\Theta}_{t-1}},\bB_{\lambda_t}} ]\]
  \[\bB_{\lambda_t} = \bzero\]

  \item Estimate the updated signals by least square (LS) algorithm
  \[
    {\bbs_t} = \arg \max_{{\bs_0}}
    {\| {\by-\sqrt {{\rho_0}M} {\mathbf{\Theta}_t}{\bs_0}}
    \|_2}
  \]

  \item Update the residual signals and the iteration number
  \[
    {\br_t} = \by-\sqrt {{\rho_0}M}
    {\mathbf{\Theta}_t}{{\overline {\bs} }_t}
  \]
  \[t = t+1\]
  \end{enumerate}

\item \emph{Output:} ${\bbs_K}$, which is the approximation most unlikely
all-zero blocks corresponding to block index set ${\Lambda_K}$ in the original
block-sparse signal vector $\bs$.
\end{enumerate}

After we get the approximation of block-sparse coefficients, we can recover
the original signals we want.

\subsection{ICBOMP Algorithm}

In addition to the block structure of the signals to recover, the amplitude of
each modulated symbol has constant modulus, which could potentially be
utilized for improved performance. More importantly, error correction coding
is usually used for correcting demodulation errors due to noise and channel
disturbances. Error detection codes such as cyclic redundancy check (CRC)
codes can be utilized to indicate whether the decoded packets are indeed
correct. Such detection cannot be perfect. However, for simplicity we will
assume CRC detection is perfect, i.e., the probabilities of miss detection and
false alarm are both zero.

Here we propose an improved algorithm ICBOMP which make use of channel coding
and CRC to carry out perfect IC in each iteration of BOMP algorithm. The idea
of IC can be found in \cite{re41,re43,re45}.

Let $\bbs_t^i$ denote each $d$-length block $\bbs_t^i$ of $\bbs_t$, which is
obtained by step 4d) in BOMP algorithm, $1 \le i \le t$. We assume that the
transmitters have used certain error correction and detection scheme for each
block signal, denoted by $[ \tLambda_t,\tbs_t ] = \cD (\bbs_t, d)$, in which
$\tbs_t$ is the output of input $\bbs_t$, and each of its blocks is denoted by
$\tbs_t^i$; $\tLambda_t$ denotes the index set for error-free blocks. For each
block, it will go through one of the two distinct operations given by function
$\cD$:

\begin{enumerate}
\item If $\bbs_t^i$ after operation of certain channel coding scheme is
error-free, then output $\tbs_t^i$ is its corrected signal vector.

\item If $\bbs_t^i$ after operation of certain channel coding scheme is not
error-free, then $\tbs_t^i = \bbs_t^i$.
\end{enumerate}

In ICBOMP algorithm, most of the calculation processes remain the same as BOMP
algorithm, and the difference occurs after signals have been updated by LS
method in step 4d) of BOMP algorithm, turning residual signals updating steps
to the following
\begin{align}
[ \tLambda_t,\tbs_t ] &= \cD (\bbs_t, d) \\
\br_t &= \by-\sqrt {\rho_0 M} \bB_{\Lambda_t}\tbs_t \\
\Lambda_t &= \Lambda_t\backslash \{ \tLambda_t \} \\
t &= t+1
\end{align}

From the above steps, we can see that, when some blocks of signals have been
exactly recovered, ICBOMP algorithm regards them as interference signals to
the following iterations and eliminates these signals, as well as their
contributions to signal receiving model of \eqref{model1}. While for the
signal blocks with errors that cannot be corrected, ICBOMP algorithm leaves
them as they were obtained through BOMP algorithm. In the case where no
error-free block is available, ICBOMP behaves the same way as BOMP.

\section{Data Recovery Guarantees}\label{Guarantees}

In this section, we will present conditions that guarantee data recovery.
Before analyzing conditions for data recovery, some notation and definitions
will be introduced first. From the definition of ${\bB}$, we can see that each
column of it is statistically normalized to one. Here we expand ${\bB}$ as
\begin{equation}
{\bB}   = [ {\underbrace {{\mathbf{b}_1} \cdots {\mathbf{b}_d}}_{\bB_1}
  \underbrace {{\mathbf{b}_{d+1}} \cdots {\mathbf{b}_{2d}}}_{\bB_2} \cdots \underbrace {{\mathbf{b}_{({N-1})d+1}} \cdots {\mathbf{b}_{Nd}}}_{\bB_N}} ]
\end{equation}

As in \cite{re23}, \cite{re24}, we give the definitions of block-coherence as
\begin{equation}
{\mu_\bB}: = \frac{1}{d}\max_{i \ne j} \| {\bB_i^H}\bB_j \|
\end{equation}
and sub-coherence as
\begin{equation}
\nu:  = \max_{1 \le l \le N} \max_{({l-1})d+1 \le i \ne j \le ld} | {\mathbf{b}_i^H{\mathbf{b}_j}} |
\end{equation}

At the same time define
\begin{equation}
{{s_l}} : = \min_{{i \in I}} {\| {\bs_i} \|_2}, \quad
{{s_u}} : = \max_{{i \in I}} {\| {\bs_i } \|_2}
\end{equation}

\subsection{Data Recovery Conditions For BOMP Algorithm} The following theorem
characterizes the block-sparse data recovery performance by BOMP algorithm.

\begin{theorem} \label{th.1}
Consider the block-sparse model in \eqref{model1}, suppose that condition
\begin{equation}   \label{th_variables}
\begin{split}
\rho_0 M[ {1-({d-1})\nu } ]^2 s_l^2 &> \tau^2
+ \rho_0 Md \mu_\bB \{ 2(N_a-1)[ 1+(d-1)\nu ]
+ N_a^2d\mu_\bB \}s_l^2\\
&\quad +2\sqrt {\rho_0 M} \tau \{ (2N_a-1)
 d \mu_\bB+[ 1+(d-1)\nu  ] \}s_l
\end{split}
\end{equation}
is satisfied, then the BOMP algorithm identifies the correct support of signal
vector $\bs$ and at the same time achieves a bounded error given by
\begin{equation}   \label{err_bound}
\| \hbs-\bs \|_2^2 \le \frac {K\tau^2} {[ {1-({d-1})\nu -({K-1})d{\mu_{\bB}}} ]^2 \rho_0 M}
\end{equation}
where $\hbs$ is the signal vector recovered by BOMP algorithm, $K \le \lfloor
\frac{MT}d \rfloor $ is the maximum number of iterations for BOMP algorithm,
$1-(d-1)\nu-(K-1)d\mu_\bB > 0$ and $\tau = \max_{1 \le j \le N} \| \bB_j^H\bz
\|_2$. For circularly symmetric complex Gaussian noise $\bz$,
\begin{equation}
P\{ \ttau \ge \| \bB_j^H\bz
\|_2\} \ge 1-e^{-\varsigma^2}\sum_{k = 0}^{d-1} \frac{(
{{\varsigma^2}})^k} {k!}
\end{equation}
where $\varsigma = \ttau /\sqrt {1+({d-1})\nu }$.
\end{theorem}

For a certain modulation constellation, suppose that each symbol's energy has
been normalized, and the minimum distance between different symbols is
${l_{\min }}$, for example, ${l_{\min }} = \sqrt 2 $ for quadrature phase
shift keying (QPSK) and ${l_{\min }} = 2$ for binary phase shift keying
(BPSK), then by the bounded error in \eqref{err_bound}, we can conclude that
the number of erroneously demodulated symbols is at most ${N_e} = \lfloor {\|
{\hbs-\bs} \|_2^2/{{({{l_{\min }}/2} )}^2}} \rfloor $. By now, we can present
the expression of symbol error rate (SER) as
\begin{equation}
  P_\text{SER} \le \frac{N_e} { {N_a}d }
\end{equation}

\remark Since ${T} > d$ and $M{T} \gg d$, we can design orthogonal columns for
precoding matrix ${\bP_n}$ of user $n$, $n = 1,2, \cdots ,N$, then each block
of dictionary ${\bB} $ is sub-matrix with orthogonal columns, meaning $\nu =
0$. On the other hand, we have $\tau \gg {{s_l}} $ when each nonzero element
of $\bs_n$ satisfies a reasonable power constrain. Additionally, if ${\mu_\bB}
= 0$, then condition \eqref{th_variables} can be simplified as
${{\rho_0}M}{{{s_l^2}}} > {\tau^2}+2\sqrt {{\rho_0}M}\tau {{s_l}} \approx
{\tau^2}$, which is milder when compared with \cite[Theorem 5]{re24}, which
yields ${{\rho_0}M}{{{s_l^2}}} > 4{\tau^2}$ when applied to our scenario.

\subsection{Conditions For Perfect IC In ICBOMP algorithm}

Thanks to the error correction and detection, ICBOMP algorithm provides better
performance than BOMP algorithm. In the case of perfect IC, the algorithm
improves signal recovery quality and also reduces computational complexity. By
eliminating the correctly decoded blocks and their contributions, the
dimensionality of useful signals is reduced, which contributes to reducing the
computations required in matrix inversion of LS algorithm. In the following,
we present a theorem that characterizes the conditions for perfect IC in each
iteration of ICBOMP algorithm.

When the first $i-1$ iterations and block selection in $i$-th iteration have
been finished, suppose $N_{ic}^{i-1}$ blocks of active users are already
correctly recovered and cancelled by the previous $i-1$ iterations, and $N_i$
blocks, whose indices are gathered to form a set $I^i$, will be substituted
into least square operation in $i$-th iteration, $N_{ic}^{i-1} + N_i=i$.
Additionally, let set $I_u^i$ contain the indices of unidentified active users
and set $I_{ic}^{i-1}$ contain the indices of active users that are already
successfully recovered and cancelled. Then the following result holds
\begin{theorem}\label{th.2}
If conditions
\begin{equation}  \label{TH2_1}
\begin{split}
\rho_0 M[ 1-(d-1)\nu_i ]^2s_{il}^2 &> \tau_i^2
 +{\rho_0}Md\mu_{i\bB}\{ 2(N_i-1)[ 1+(d-1)\nu_i ]
 +N_i^2d\mu_{i\bB} \}s_{il}^2\\
&\quad+2\sqrt {\rho_0 M} \tau_i\{ (2N_i-1)d\mu_{i\bB}
 +[ 1+(d-1)\nu_i ] \}s_{il}
\end{split}
\end{equation}
and
\begin{equation}   \label{TH2_2}
[ 1-(d-1)\nu_i -(N_i-1)d\mu_{i\bB} ]^2\rho_0 M t_c l_{\min }^2
  \ge 4\tau_i^2
\end{equation}
are satisfied, then at least one block will been successfully recovered and
cancelled in $i$-th iteration of ICBOMP algorithm. In \eqref{TH2_1} and
\eqref{TH2_2}, $t_c$ is the number of bits that can be corrected by the
channel coding scheme,
\begin{equation}
\tau_i = \max_{j \in \{ [ N ]\backslash \{ I_{IC}^{i-1}
  \cup I_c^i \} \}}
  {\| \bB_j^H(\sqrt {\rho_0 M} \bB_{I_c^i}\bs_{I_c^i}+\bz) \|_2}
\end{equation}
and $\mu_{i\bB}$, $\nu_i$ and $s_{il}$ are respectively defined as $\mu
_{\bB}$, $\nu $ and $s_l$, and their definitions are only limited to users or
active users in $\{ [ N ]\backslash \{ I_{ic}^{i-1} \cup I_u^i \} \}$.
\end{theorem}

In \thref{th.2}, we only considered the case where all blocks in $I^i$
correspond to active users, based on the consideration that if signals of an
active user cannot be successfully recovered when all the active users have
been identified, they are less likely to be successfully recovered when
non-active blocks begin to enter into the least square algorithm.

\begin{IEEEproof}[Proof of Theorem 2]
\thref{th.2} can be viewed as a corollary of \thref{th.1}.

When we consider each iteration in ICBOMP algorithm, by \eqref{TH2_1} which
yields \eqref{okay_0} in Appendix A, when applied to $i$-th iteration, the
most likely active users will be identified. Treat the users in $I_u^i$ and
noise as perturbations for recovering the signal of users in $I^i$. With the
same proof for \thref{th.1}, \eqref{TH2_1} can be verified. When \eqref{TH2_1}
is satisfied, we achieve an error bounded by
\begin{equation}
\| \hbs_{I^i}-\bs_{I^i} \|_2^2 \le \frac{N_i\tau_i^2}{[ 1-(d-1)\nu_i -(N_i-1)d\mu_{i\bB} ]^2\rho_0 M}
\end{equation}
where we have utilized the knowledge $| I^i | = N_i$ which can be exactly
obtained by ICBOMP algorithm. If condition
\begin{equation}   \label{TH2_3}
\| \hs_{I^i}-s_{I^i} \|_2^2 \le N_it_c(\frac{l_{\min }}{2})^2
\end{equation}
holds, then by correction of channel coding, at least one user's signals will
surely be recovered without an error. By \eqref{TH2_3}, \eqref{TH2_2} is
obtained.
\end{IEEEproof}

\subsection{Condition From Information Theoretic Point Of View} From the BS's
point of view, it is desirable to recover all the information conveyed by
$\bs$, including number of active users, exact indices of these active users,
their transmitted information bits, etc.. When all the information are
measured by bits, then The number of bits representing the indices of active
users and signal bits of the transmitted messages are respectively $\log_2
\binom{N}{N_a}$ and $\sum_{i = 1}^{N_a} b_i$. Assume all bits are generated
with equal probability, and let $S$ denote the set of bits needed to represent
the total information, then

\begin{equation} \label{eq.ss}
| S | \ge \log_2 \binom{N}{N_a}+\sum_{i = 1}^{N_a} b_i
\end{equation}

\remark When the number of active users and lengths of the messages of active
users are not prior known to BS, then the inequality in \eqref{eq.ss} is
strictly established. Even when these two factors are prior known to BS,
\eqref{eq.ss} still holds.

The following theorem roughly shows that the total bits of all information
that can be recovered at the receiver can not exceed the capability of channel
in a frame time.

\begin{theorem} \label{th.3}
Define $p_e$ as the probability that some error has happened in the recovery
of the information in set $S$, Then the following condition is necessary for
the data recovery
\begin{equation}   \label{IT}
| S | \le  \frac 1 {1-p_e} [H(p_e)+\log_2\det (\bI_{MT}+\rho
_0\bB_I\bB_I^H)]
\end{equation}
\end{theorem}

\begin{IEEEproof}[Proof of Theorem 3]
Our proof of \thref{th.3} mainly includes the properties of entropy, mutual
information and Fano's Inequality \cite{re44}.

We have
\begin{align}
I(S;\bY,\bB) &= I(\bs;\bY,\bB)\\
 &= I(\bs;\bB)+I(\bs;\bY|\bB)\\
 &= I(\bs;\bY|\bB)
\end{align}
in which independence between $\bs$ and ${\bB}$ are utilized, which means
$I(\bs;\bB)=0$. By property
\begin{align}   \label{H_s}
H(S|\bY,\bB) &= H(S)-I(S;\bY,\bB)\\
&= H(S)-I(\bs;\bY|\bB)\\
&\ge | S |-C
\end{align}
where $C =\max_{p(\bX)} I(\bs;\bY|\bB)$ is the maximum mutual information
(channel capacity). On the other hand, by Fano's Inequality
\begin{equation}   \label{F_I}
H(S|\bY,\bB) \le H(p_e)+p_e| S |
\end{equation}
combining \eqref{H_s} and \eqref{F_I}, and using the well-known result $C =
\log_2\det (\bI_{MT}+\rho_0\bB_I\bB_I^H)$ \cite{re28}, the desired inequality
follows.
\end{IEEEproof}
\remark It can be seen that when $p_e\to 0$, the right hand side of the
inequality converges to $C$. It means that we cannot hope to decode correctly
information (including all information useful to the BS) at a rate higher than
the capacity of the channel, assuming the availability of the information of
the set of active users and their channels.

\section{Numerical Results}\label{simulations}

The experimental studies for verifying the proposed scheme are presented in
this section. In all simulations, the channel response matrix is i.i.d.
Gaussian matrix of complex values and the $N_a$ active users are chosen
uniformly at random among all $N$ online users. As for the block-sparse data
vectors to be transmitted, we assume QPSK for data modulation. All results are
presented with symbol error rate (SER) and frame error rate (FER) versus
$E_s/N_0$, where $E_s$ is the symbol energy, $N_0$ is the noise spectral
density. In the simulations with BOMP algorithm, we do not set the number of
antennas to a large value, say one hundred or more, for the sake of
simplicity. Besides, we will choose the frame length to be a multiple of the
maximum length of short messages.

\subsection{Influences Of Different Parameters}

\begin{figure}[ht]
\centering
\includegraphics[width=\figwidth]{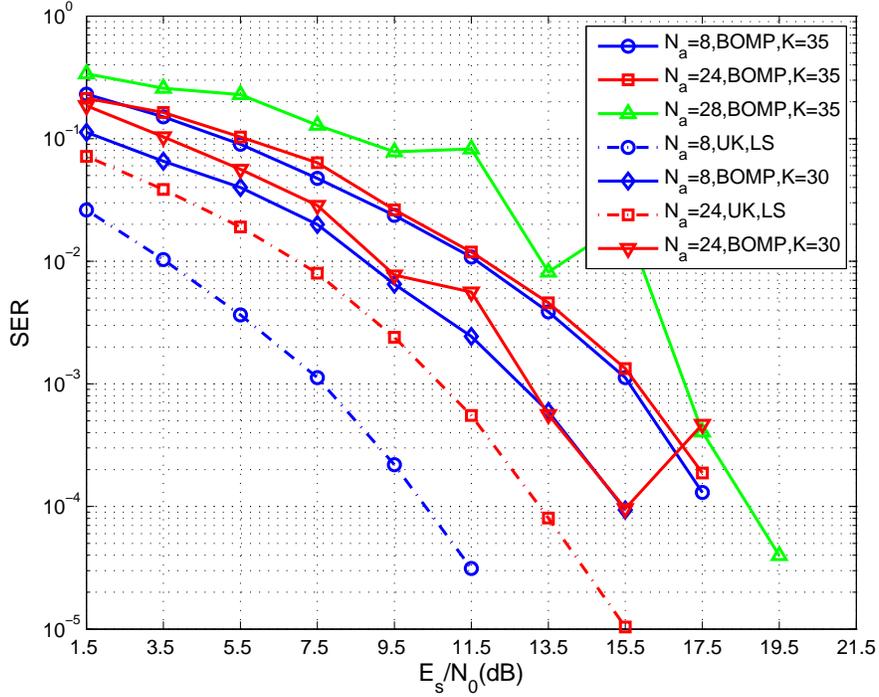}
\caption{symbol recovery with 8 antennas at BS}
\label{M8users3}
\end{figure}

\begin{figure}[ht]
\centering
\includegraphics[width=\figwidth]{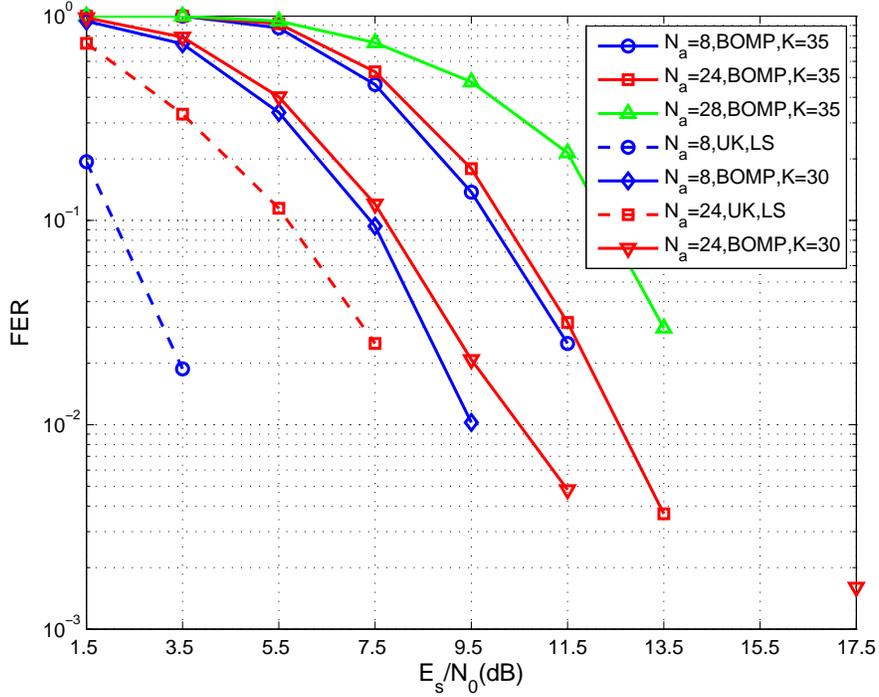}
\caption{frame recovery with 8 antennas at BS}
\label{M8users3_FER}
\end{figure}

\begin{figure}[ht]
\centering
\includegraphics[width=\figwidth]{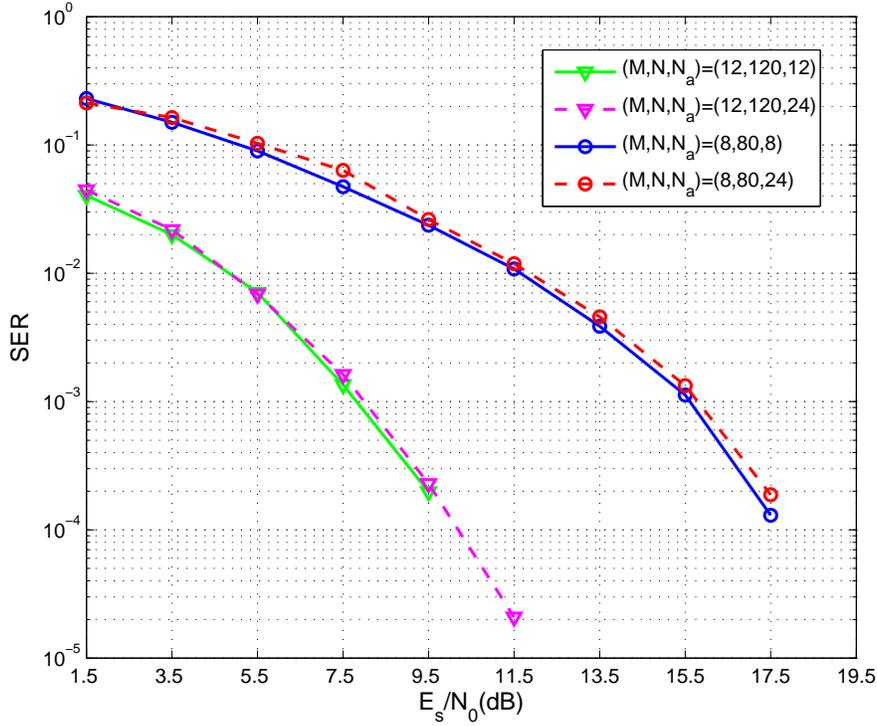}
\caption{symbol recovery with different numbers of antennas at BS}
\label{antenna}
\end{figure}

\begin{figure}[ht]
\centering
\includegraphics[width=\figwidth]{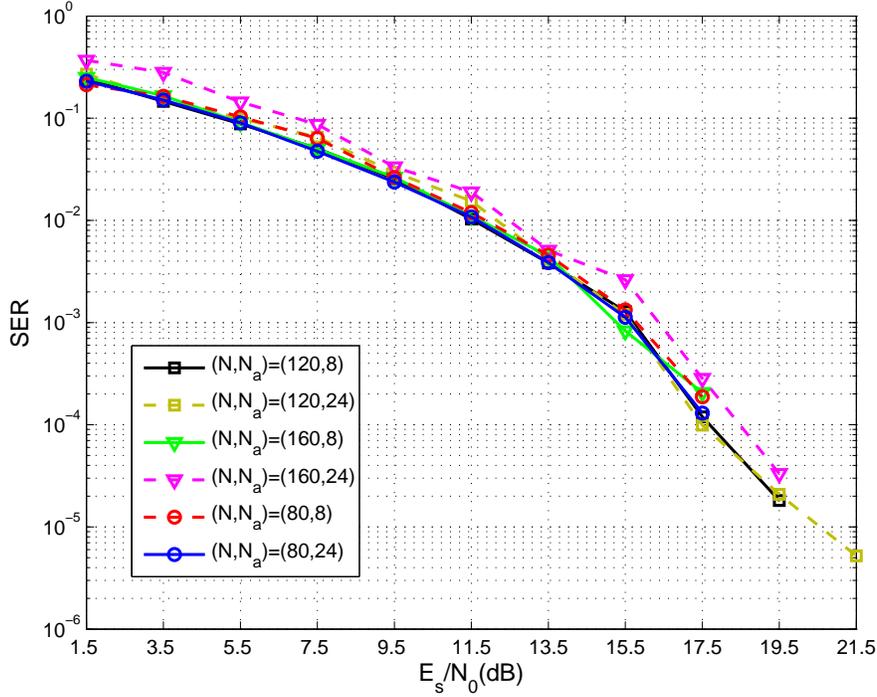}
\caption{symbol recovery with different numbers of online users}
\label{potential_users3}
\end{figure}

\begin{figure}[ht]
\centering
\includegraphics[width=\figwidth]{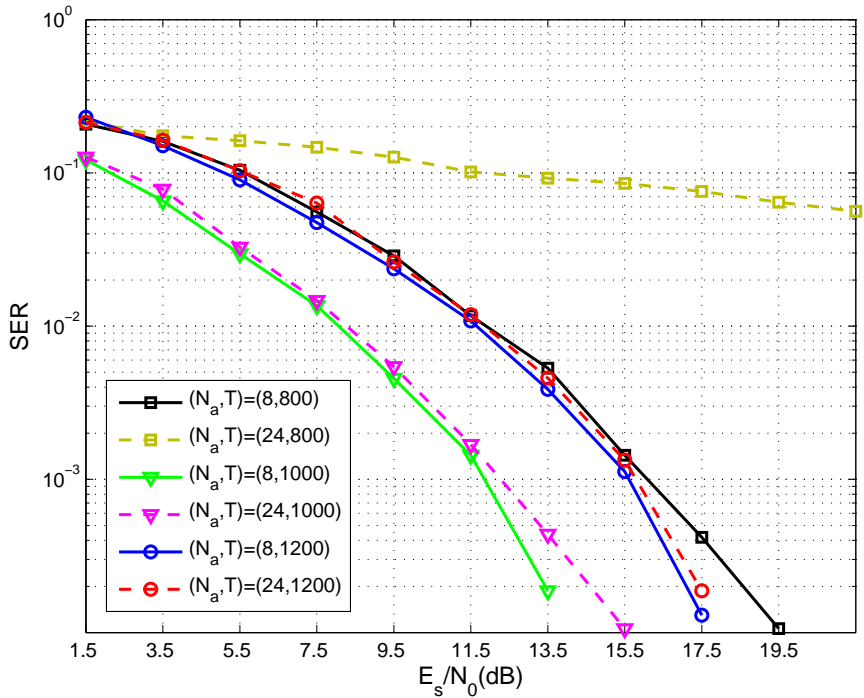}
\caption{symbol recovery with different lengths of frames}
\label{spreadT}
\end{figure}

\begin{figure}[ht]
\centering
\includegraphics[width=\figwidth]{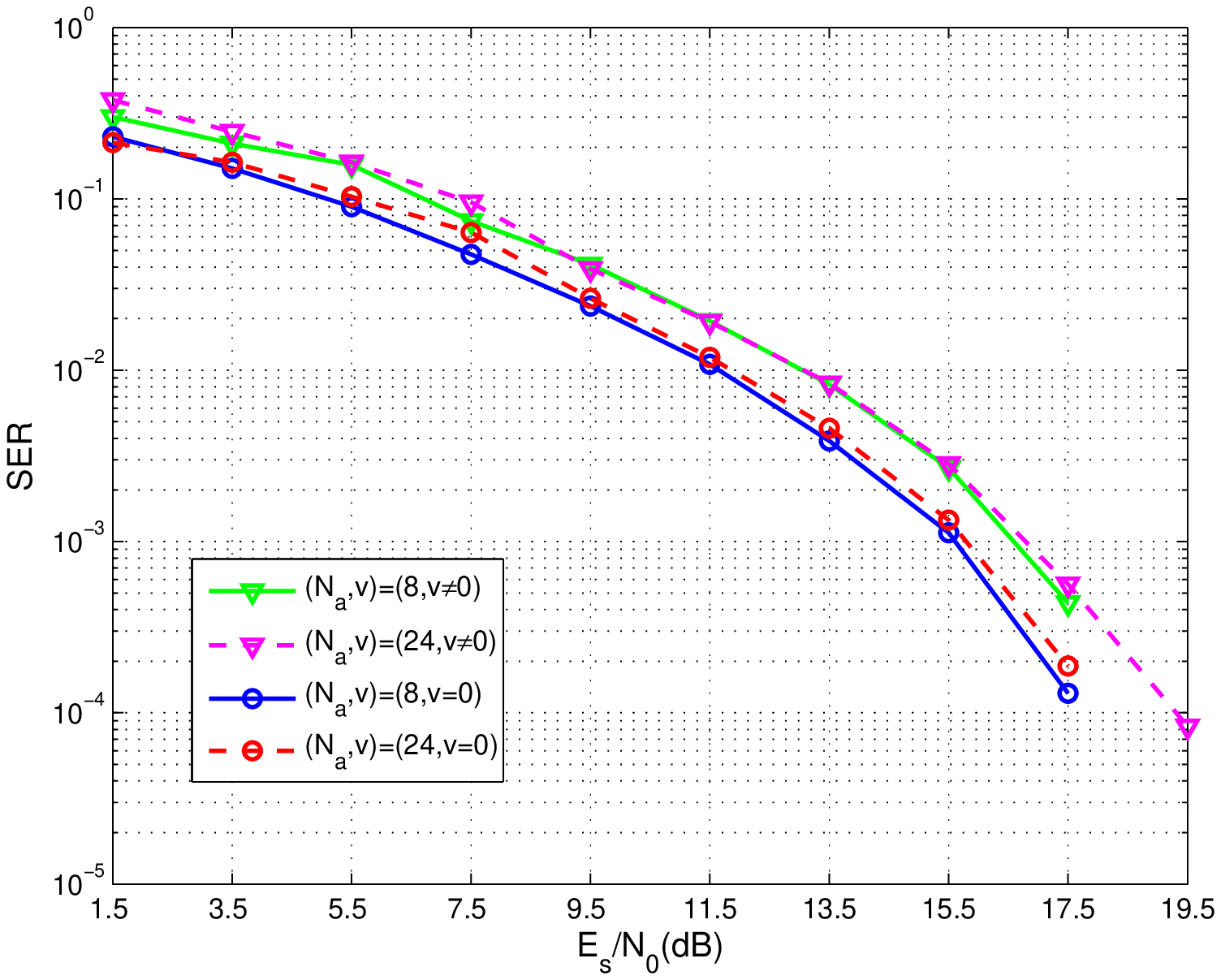}
\caption{symbol recovery with or without column orthogonal blocks}
\label{orthogonal}
\end{figure}

In the first experiments, we have the simulations of BOMP algorithm to check
the influences of different parameters. We assume that all the messages have
the same length $d$, and we simply design $\bP_n$ a random matrix with $(v =
0)$ or without $(v \ne 0)$ orthogonal columns, $n = 1,2, \ldots ,N$.

In our simulation, the SER is computed as follows: when a demodulated signal
of an active user is different from its original signal, we claim a symbol
error; if an active user is not identified, then all $d$ symbol of that user
are treated as erroneous.

\testcase \figref{M8users3} shows the performance of the proposed scheme with
8 antennas at BS, where $K$ is the number of iterations for BOMP algorithm.
Other parameters are given as $(N,d,T,v) = (80,200,1000, 0)$. The results
indicate that, the SER increases when the number of active users becomes
larger. For case where number of iterations is 35, when the number of active
users is lower than a certain number, say 24 in our results, the SER is
basically independent of the number of active users. Besides, we have observed
that out of 35 iterations, in most cases the $N_a$ ($N_a < 24$) active users
can be successfully identified.

Also in \figref{M8users3}, we give results when less number of iterations is
set for BOMP algorithm. When there are not too many active users, such as 8
users, fewer iterations are needed. But for 24 active users, 30 iterations are
not enough to include all the active users. We also plotted two curves of
reference in dotted lines. Both curves were obtained under the conditions that
the actual active users are already known (users known, UK) and picked out.
From the results, we can conclude that the number of non-active users has a
great influence on the performance, even greater than that of active users.

\testcase Corresponding to \figref{M8users3} in SER, the FER is depicted
\figref{M8users3_FER} with the same settings. In our simulation, the FER is
computed as follows: when more than 8 bits in a message are demodulated in
error, we claim a frame error. If the bit errors are equal to or less than 8,
we hypothesize that they can be detected and corrected by the channel coding
schemes. The same trend in FER performance can be observed as SER. When the
$E_s/N_0$ exceeds a certain threshold, the FER will be negligible.

The normalized throughput is defined as $(1-P_\text{FER})N_ad/(MT)$, where
$(1-P_\text{FER})N_a$ is the maximum number of allowed active users in our
scheme, $P_\text{FER}$ is the value of FER; and $MT/d$ is the maximum number
of users that can be served when all time slots of a frame are effectively
used for data transmission, which is $40$ under the given parameters. If 24
active users are allowed to be simultaneously served, the throughput will
reach 60\% of maximum possible. In contrast, in conventional random access
protocols, if we treat the signaling messages (such as request-to-send (RTS)
signaling \cite{re6}, \cite{re7}) also as small packets, the system throughput
will be no more than 60\%. Furthermore, if collision happens, which is often
the case, the throughput will decrease a step further. Therefore, our scheme
will greatly improve the system throughput compared to conventional schemes.

\testcase In \figref{antenna}, we compare the performance when BS are equipped
with different numbers of antennas. Other parameters are given as $(d,T,K,v) =
(200,1000,35,0)$. The results show that, when the number of antennas $M$
increases, the SER performance becomes remarkably better and a higher ratio
$N_a/M$ can be accepted. On the other hand, a big performance gap between 8
antennas and 12 antennas at BS is observed. More antennas at BS allows a
larger number of iterations for BOMP algorithm to accommodate more active
users, and the big performance gap appears when we set both cases the same
number 35 of iterations.

\testcase From \figref{potential_users3} with parameters
$(M,d,T,K,v)=(8,200,1000,35,0)$, we can see that when the number of active
users is fixed, the SER increases as the number of online users increases, but
the performance degradation is rather small, even when the number of online
users has been doubled, nearly no more than 1dB degradation can be observed
for 24 active users. By \thref{th.3}, the number of online users is not the
dominant factor to affect the performance under the given parameters.

\testcase \figref{spreadT} depicts the performance when frame lengths are
different, respectively for $T=4d$, $T=5d$ and $T=6d$. Other parameters are
given as $(M,N,d,v)=(8,80,200,0)$. The number of iterations $K$ is set to 28,
35 and 42, respectively. The results show that the longer the frame length is,
the better performance, and hence the more users that can be simultaneously
served. However, affected by the normalization of columns in precoding matrix,
even when the length of frame grows, the benefits diminish. This phenomenon
will be observed when parameters are chosen to ensure that $MT/(dK)$ is a
constant.

\testcase In \figref{orthogonal}, we investigate the effect of orthogonality
of the block of the dictionary on the performance. Other parameters are given
as $(M,N,d,T,K)=(8,80,200,1000,35)$. As we can see, for column non-orthogonal
case, nearly 1dB performance degradation is observed when compared with that
of column orthogonal condition.

\begin{table}
\caption{Theoretical Analysis ($v = 0$)} \label{table_ex} \centering
\newcommand{\bb}[1]{\raisebox{-2ex}[0pt][0pt]{\shortstack{#1}}}
\begin{tabular}{ccccccccccc}\hline
\bb{$M$}&\bb{$N$}&\bb{$d$}&\bb{$T$}&\bb{$\langle {{s_l}} \rangle(\langle
{{s_u}}
\rangle)$}&\bb{$\beta$}&\bb{$\mu_{\bB}$}&\bb{$\tau$}&\multicolumn{3}{c}{$N_{a.lowwer}(E_s/N_0)$}\\
\cline{9-11} &&&&&&&&0dB&10dB&15dB\\\hline
8&80&200&1000&$14.14$&2&$0.0035$&$15.00$&0&1&1\\
50&500&200&1000&$14.14$&2&$0.0019$&$15.00$&1&1&1\\
100&1000&200&1000&$14.14$&2&$0.0014$&$15.00$&1&2&2\\\hline
8&80&200&1000&$14.14$&2&$0.0035$&$14.20$&0&1&1\\
50&500&200&1000&$14.14$&2&$0.0019$&$14.20$&1&1&1\\
100&1000&200&1000&$14.14$&2&$0.0014$&$14.20$&1&2&2\\\hline
8&80&100&500&10&2&$0.0066$&$15.00$&0&1&1\\
50&500&100&500&10&2&$0.0037$&$15.00$&1&1&1\\
100&1000&100&500&10&2&$0.0030$&$15.00$&1&1&1\\\hline
\end{tabular}
\end{table}

The results of our theoretical analysis of \thref{th.1} for the case where $v
= 0$ and all messages have same length are shown in Table \ref{table_ex},
giving minimum number of active users that can be simultaneously served.
Obviously, the bounded minimum number of active users is pessimistic when
compared with our simulation results. At the same time, such pessimistic
result can also be seen in the theoretical analysis for SER, for condition
$1-(d-1)\nu -(K-1)d\mu_\bB > 0$ can not always be satisfied.

\remark In all above simulations, we have set $d$ as the length of all
messages. In practice, this may not be the case. In fact, when different
lengths for messages exist and the number of active users is large, it has
some slight performance degradation. For example, similar result can be
observed when there are 24 active users and length of each message is
uniformly distributed in the interval from 50 to 200. On the other hand,
performance gap between $d=100$ and $d=200$ is barely visible.

\subsection{Performance Of ICBOMP Algorithm}

In this part, we will apply ICBOMP algorithm to recover the transmitted
signals.

\begin{figure}[ht]
\centering
\includegraphics[width=\figwidth]{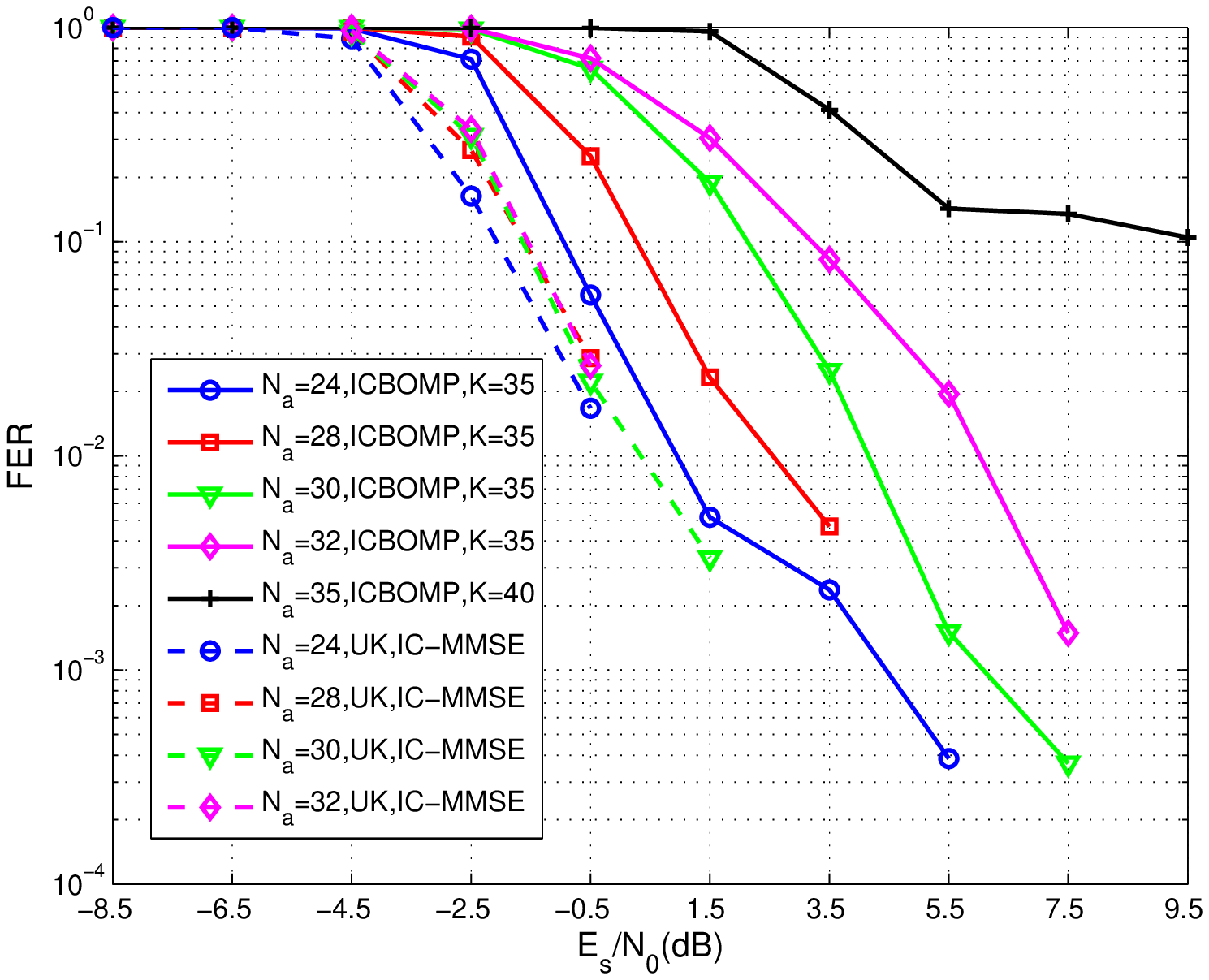}
\caption{Frame recovery with 8 antennas at BS by ICBOMP algorithm}
\label{ImprovedBOMP3}
\end{figure}

\testcase The results of FER for 8 antennas at BS are depicted in
\figref{ImprovedBOMP3}, and other parameters are given as
$(N,d,T,v)=(80,200,1000,0)$. For ICBOMP algorithm, we choose the channel
coding scheme that is capable of correcting at most 8 bits for message of 200
symbols, e.g., a shortened BCH(472,400), which enjoys a relatively high code
rate. When the error of a message is beyond correction, a frame error is
declared.

\begin{figure}[ht]
\centering
\includegraphics[width=\figwidth]{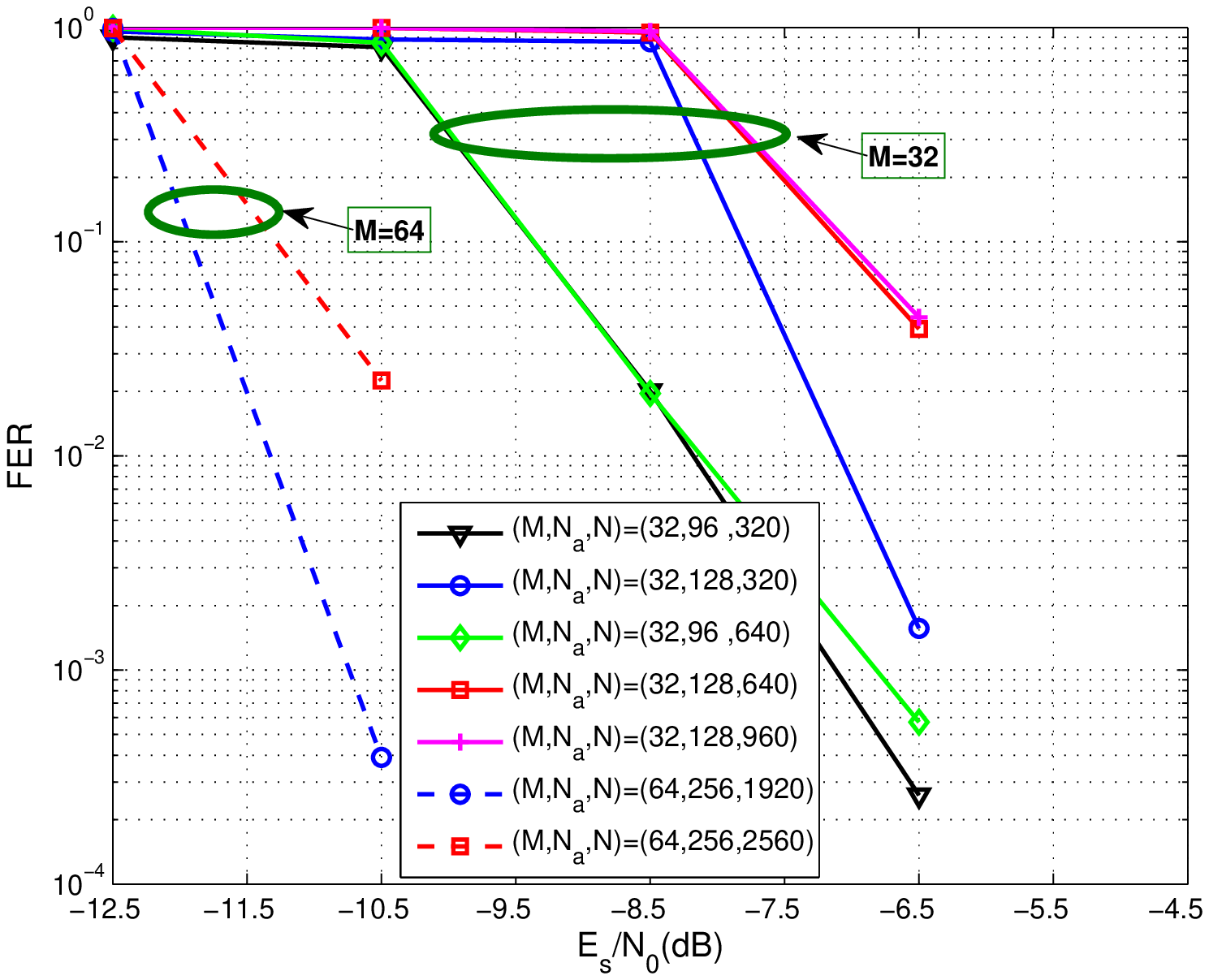}
\caption{frame recovery with 32 and 64 antennas at BS by ICBOMP algorithm}
\label{M32_64}
\end{figure}

Compared with \figref{M8users3} achieved by BOMP algorithm,
\figref{ImprovedBOMP3} shows that ICBOMP algorithm can greatly improve the
recovering performance, and more active users can be served simultaneously,
say 30 or 32. Thus the throughput will be increased a step further, reaching
75\% or 80\% of the maximum possible. When ICBOMP algorithm is applied, the
result we obtain for 24 active users with 35 iterations is almost identical to
that achieved by 30 iterations, and we just depict one of them for clarity.
Therefore, ICBOMP algorithm can also narrow the performance gap between
different iteration numbers. It is also noticed that the performance gap
between different numbers of active users have been widened.

Also, we included several curves in dotted lines to show the performance when
actual active users are already picked out and no non-active users are chosen
and interference-cancellation minimum mean-squared error (IC-MMSE) receiver is
used. The IC-MMSE receiver performs iteratively MMSE decoding and perfect IC
until more iterations no longer benefit. Perfect IC in IC-MMSE receiver is
operated as in the ICBOMP algorithm. It shows that our ICBOMP receiver
achieves a little worse performance than the IC-MMSE receiver, about 1dB
performance degradation for 24 active users and 2dB performance degradation
for 28 active users. Although for ICBOMP receiver, performance degradation
exists when compared with IC-MMSE receiver, it is still highly competitive,
since it requires no knowledge about active users. In fact, the ICBOMP
receiver is able to achieve better performance than MMSE receiver when the
number of active users is no more than 30.

\testcase For massive MIMO, 8 or 12 antennas at BS are not enough. In the
following, we present the FER performance with 32 and 64 antennas at BS in
\figref{M32_64}, and other parameters are given as $(d,T,v) = (31,155,0 )$. To
ease the computational burden, we set the length for each message to 31.
Channel coding scheme is assumed such that up to 1 bit of error can be
corrected; e.g., a shortened BCH(69,62) could be used. When the number of
errors in a message is larger than the designed error correction capability, a
frame error is declared. The results show that, with more antennas, great
improvements in performance are observed, and many more active users and
online users can be accommodated.

\section{Discussions}\label{Discussion}

In this section, we discuss a few issues related to the design of our proposed
scheme and its practical significance.

\subsection{Dictionary Design}

By \thref{th.1}, the smaller the block-coherence $\mu_\bB$ and $\nu$ are, the
better performance we can achieve, and thus the more active users the model
can simultaneously accommodate. The authors in \cite{re23} had a discussion
about the design of the dictionary that can lead to significant improvement in
data recovery in block-sparse model. In our model, only the precoding matrix
is up to our design.

In our scheme, when the number of active users is more than the number of
antennas at BS, the channel vectors among users are correlated, even with
massive MIMO technique. However, by our precoding scheme, correlations among
columns in $\bB$ can be smaller than correlations among channel vectors of
different users, which means that block-coherence $\mu_\bB$ can still be
rather small, as long as precoding is well designed.

\subsection{Application To Asynchronous Setting}

Random MAC channel is usually asynchronous \cite{re35}, while in our model,
frame synchronization is assumed. In fact, our model can also be adjusted to
quasi-synchronous setting, as long as the maximum asynchronism is known to the
receiver. In such scenario, we can handle the problem by lengthening the
maximum length $d$ with addition of the maximum asynchronous level, leading to
a slight adjustment in the length of each precoding sequence. Readers can
refer to \cite{re35} for more details.

\subsection{Sparsity Level Selection} In our model, since we do not know
exactly the number of active users a priori, a large number of iterations are
usually needed for BOMP or ICBOMP algorithm. When there are not too many
active users, however, unnecessary iterations increases computational cost.
For example, our statistical results indicate that, when $E_s/N_0 \ge 10$dB
and with the same conditions for simulation of $8$ active users in
\figref{M8users3}, 12 iterations for BOMP algorithm will correctly identify
all the active users with a probability exceeding 99.9\%. And for ICBOMP
algorithm, even less iterations and higher probability can be anticipated. To
address this problem, some methods for sparsity level selection will work,
such as sparsity adaptive matching pursuit in \cite{re40}. When the number of
active users is large, sparsity level identification may not bring too much
benefit.

\subsection{Message Segmentation} Messages of small packet can be segmented
into shorter parts further, and each part can be transmitted with our scheme.
When this method is adopted, we not only alleviate the computational
complexity, but also ease the requirement for coherent time of channel.
Besides, we can decrease the length uncertainty for each segmented part, and
shorten the transmission duration for small packets whose lengths are short.
More importantly, such approach may even be adopted to data transmission for
messages that are not belonging to small packet.

\section{Conclusion}\label{Conclusion}

In this paper, we proposed an uplink data transmission scheme for small
packets. The proposed scheme combines the techniques of block precoding and
sparsity-aware detection. It is especially suitable for system with a large
number of antennas at the base station. Under the assumption that the BS has
perfect CSI of every online user, we developed a block-sparse system model and
adopted BOMP algorithm and its improved version ICBOMP algorithm to recover
the transmitted data. With the ICBOMP algorithm which utilizes the function of
error correction and detection coding to perform perfect IC in each iteration,
an significant performance improvement was observed.

The transmission scheme considered in this paper is applicable to future
wireless communication system. The reason is that small packets play a more
and more important role in the data traffic due to the wide usage of
intelligent terminals. The overall throughput of such a system is currently
hampered by small packets because of the heavy signaling overhead. Our scheme
will greatly reduce the signaling overhead and improve the throughput of such
systems.

\appendices
\section{Proof of \thref{th.1}} We first present a few results and lemmas that
are useful for the proof of \thref{th.1}. Our proof of \thref{th.1} follows
along the lines of \cite{re24}.

First of all, we present two useful results that have been obtained in some
literature cites.
\begin{result} \label{lem.coh}
\cite[Lemma 1]{re24} Given the dictionary $\bB $ of normalized columns with
block-coherence $\mu_\bB$ and sub-coherence $\nu$, it holds that

\begin{equation} \label{lem.coh1}
\max_{i \ne j} \| \bB_i^H\bB_j \| \le d\mu_\bB
\end{equation}
and
\begin{equation} \label{lem.coh2}
1-(d-1)\nu \le \| \bB_i^H\bB_i \| \le 1+(d-1)\nu
\end{equation}

Provided that $1-(d-1)\nu-(K-1)d\mu_\bB > 0$ and $| I | \le K$, it holds that
\begin{equation} \label{lem.coh3}
\| (\bB_I^H\bB_I)^{-1} \| \le [ 1-(d-1)\nu -(K-1)d\mu_\bB ]^{-1}
\end{equation}
\end{result}

\begin{result} \label{lem.trace}
\cite[\textsection 10.2]{re32} Denote $m \times n$ matrices $\bA$ and $\bC$,
whose singular values are $\sigma_1 \ge \ldots \ge \sigma_n$, $\gamma_1 \ge
\ldots \ge \gamma _n$, respectively, then
\begin{equation}
-\sum_{i = 1}^n \sigma_i\gamma_i  \le \Re [ \trace(\bA\bC^H) ] \le \sum_{i = 1}^n \sigma_i\gamma_i
\end{equation}
\end{result}

In the following, two lemmas will be given.

\begin{lemma}\label{lem.bs}
Consider the block-sparse model in model \eqref{model1} and the condition
$\tau = \max_{1 \le j \le N} \| \bB_j^H\bz \|_2$. Provided that
\begin{equation}  \label{okay_0}
\begin{split}
\rho_0 M[1-(d-1)\nu ]^2s_u^2 &> \tau^2+\rho_0 M (d\mu_\bB)^2N_a^2s_u^2
  +2\rho_0 Md\mu_\bB\{ (N_a-1)[ 1+(d-1)\nu  ]  \}s_l^2\\
& \quad +2\sqrt {\rho_0 M} \tau  \{ (N_a-1)d\mu_\bB+[ 1+(d-1)\nu  ] \}s_l
 +2\sqrt {\rho_0 M}N_ad\mu_\bB \tau s_u
\end{split}
\end{equation}
it holds that
\begin{equation}
\max_{j \in I} \| \bB_j^H\by \|_2 > \max_{j \notin I} \| \bB_j^H\by \|_2
\end{equation}
\end{lemma}

\begin{IEEEproof}[Proof of \lemref{lem.bs}]
Note that
\begin{equation}
\begin{split}
\| \bB_j^H \by \|_2^2 &= \by^H\bB_j\bB_j^H\by\\
&= \trace\{ \bB_j^H\by\by^H\bB_j \}\\
&= \trace\left\{ \bB_j^H\left[ \sqrt {\rho_0 M} (\sum_{i \in I} \bB_i\bs_i
)+\bz \right]  \left[ \sqrt {\rho_0 M} (\sum_{i \in I} \bB_i\bs_i )+\bz
\right]^H\bB_j \right \}
\end{split}
\end{equation}

Then we have
\begin{equation}
\begin{split}
\max_{j \notin I} \| \bB_j^H\by
\|_2^2 &= \max_{j \notin I} \trace\left( \bB_j^H\by\by^H\bB_j \right)\\
&= \max_{j \notin I} \rho_0 M \cdot \trace\left[ (\sum_{i \in I}
  \bB_j^H\bB_i\bs_i )(\sum_{i \in I} \bs_i^H{\bB_i^H}\bB_j ) \right]\\
&\quad +\max_{j \notin I} 2\sqrt {\rho_0 M}  \cdot
  \Re \left\{ \trace\left[(\sum_{i \in I} \bB_j^H\bB_i\bs_i )\bz^H\bB_j
  \right] \right\}
+\max_{j \notin I} \trace\left( \bB_j^H\bz\bz^H\bB \right)
\end{split}
\end{equation}

Note that for vectors $\tbx \in \mathbb{C}^{n \times 1}$ and $\tby \in
\mathbb{C}^{n \times 1}$, the matrix $\tbx\tby^H$ at most has one nonzero
singular value which equals to the absolute value of $\tbx^H\tby$. In
addition, for any matrix $\bA \in \mathbb{C}^{n \times n}$, matrix
$\tbx\tby^H\bA$ also has at most one nonzero singular value which equals to
the absolute value of $\tbx^H\bA^H\tby$, and it holds that $\| \tbx\tby^H\bA
\|= | \tbx^H\bA^H\tby | = | (\tbx,\bA^H\tby) | \le \| \tbx \|_2\| \bA^H\tby
\|_2 \le \| \bA \|\| \tbx \|_2\| \tby \|_2$. Together with \reref{lem.bs} and
\reref{lem.trace}, we have
\begin{align}
\max_{j \notin I}& \trace\{ (\sum_{i \in I} \bB_j^H\bB_i\bs_i )
  (\sum_{i \in I} \bs_i^H\bB_i^H\bB_j ) \} \\
&\le N_a\{\max_{i\ne j}
  \trace\{(\bB_j^H\bB_i)(\bs_i\bs_i^H\bB_i^H\bB_j) \}\}
 +(N_a^2-N_a)\{ \max_{i \ne j \ne r} \trace\{ (\bB_j^H\bB_i)
  (\bs_i\bs_r^H\bB_r^H\bB_j) \} \}\notag\\
 &\le N_a\{ \max_{i \ne j} \| \bB_j^H\bB_i\| \|
  \bs_i\bs_i^H\bB_i^H\bB_j\| \} +(N_a^2-N_a)\{ \max_{i \ne j \ne r} \|
  \bB_j^H\bB_i\| \| \bs_i\bs_r^H\bB_r^H\bB_j\| \}
\end{align}
\begin{align}
 &\le N_a\{ d\mu_\bB d\mu_\bB s_u^2 \}+(N_a^2-N_a)\{ d{\mu_\bB}d\mu
_\bB s_u^2 \}\\
&=N_a^2(d\mu_\bB)^2s_u^2
\end{align}
where we have used the identity $\Re [ \trace(\bA\bA^H) ] = \trace(\bA\bA^H)$.
Furthermore, we have
\begin{align}
\max_{j \notin I} \Re\left \{ \trace\left[ (\sum_{i \in I} \bB_j^H\bB_i\bs_i
)\bz^H\bB_j \right] \right \} & \le N_a \max_{i \ne j}
\Re [ \trace( \bB_j^H\bB_i\bs_i\bz^H\bB_j ) ] \\
 & \le N_a \max_{i \ne j} \| \bB_j^H\bB_i\| \| \bs_i\bz^H\bB_j\| \\
& =N_ad\mu_\bB\tau s_u
\end{align}

By the definition of $\tau$
\begin{equation}
\max_{j \notin I} \trace\{ \bB_j^H\bz\bz^H\bB_j \} = \max_{j \notin I} \| \bB_j^H\bz \|_2^2 \le \tau^2
\end{equation}

From derivations above, we obtain
\begin{equation}
\max_{j \notin I} \| \bB_j^H\by \|_2^2 \le \rho_0 MN_a^2(d\mu_\bB)^2 s_u^2 +2\sqrt {\rho_0 M} N_ad\mu_\bB\tau  s_u +\tau^2
\end{equation}

On the other hand, we have
\begin{align}
\max_{j \in I} \trace\{ (\sum_{i \in I} \bB_j^H\bB_i\bs_i )(\sum_{i \in I} \bs_i^H\bB_i^H\bB_j ) \}
& = \max_{j \in I} \trace\{ \bB_j^H\bB_j\bs_j\bs_j^H\bB_j^H\bB_j \}\\
& \quad +\max_{j \in I} 2\Re \{ \trace\{ (\bB_j^H\bB_j\bs_j)(\sum_{i \in I\backslash \{ j \}} \bs_i^H{\bB_i^H}\bB_j ) \} \}\\
& \quad +\max_{j \in I} \trace\{ (\sum_{i \in I\backslash \{ j \}} \bB_j^H\bB_i\bs_i )(\sum_{i \in I\backslash \{ j \}} \bs_i^H\bB_i^H\bB_j ) \}
\end{align}

For each summation term above, the first term
\begin{align}
\max_{j \in I} \trace\{ \bB_j^H\bB_j\bs_j\bs_j^H\bB_j^H\bB_j \}
& =\max_{j \in I} \{ \bs_j^H(\bB_j^H\bB_j)^2\bs_j \}\\
 & \ge \lambda_{\min }\{ (\bB_j^H\bB_j)^2 \}\max_{j \in I} \{ \| \bs_j\bs_j^H \| \}\\
& =[ 1-(d-1)\nu  ]^2s_u^2
\end{align}
in which Gershgorin circle theorem and Rayleigh-Ritz theorem have been used.
By Gershgorin circle theorem and the property of eigenvalue, all the
eigenvalues of $(\bB_j^H\bB_j)^2$ are in the range \\ $[ (1 - (d-1)\nu
)^2,(1+(d-1)\nu )^2 ]$. At the same time, just like the derivation when $j
\notin I$, the second summation term can be bounded as
\begin{align}
\max_{j \in I} 2\Re \{ \trace\{ (\bB_j^H\bB_j\bs_j)(\sum_{i \in I\backslash \{ j \}} \bs_i^H\bB_i^H\bB_j ) \} \}
 \ge -2(N_a-1)[ 1+(d-1)\nu  ]d\mu_\bB s_l^2
\end{align}
And for the third term
\begin{equation}
\max_{j \in I} \trace\{ (\sum_{i \in I\backslash \{ j \}} \bB_j^H\bB_i\bs_i )(\sum_{i \in I\backslash \{ j \}} \bs_i^H\bB_i^H\bB_j ) \} \ge 0
\end{equation}

Besides, since
\begin{align}
\max_{j \in I}& \Re \{ \trace\{ (\sum_{i \in I} \bB_j^H\bB_i\bs_i )\bz^H\bB_j \} \}\\
& =\max_{j \in I} \Re \{ \trace\{ \sum_{i \in I\backslash \{ j \}} \bB_j^H\bB_i\bs_i\bz^H\bB_j +\bB_j^H\bB_j\bs_j\bz^H\bB_j \} \}\\
 & \ge -(N_a-1)d\mu_\bB\tau s_l -[ 1+(d-1)\nu  ]\tau s_l
\intertext{and}
\max_{j \in I}& \trace\{ \bB_j^H\bz\bz^H\bB_j \} \ge 0
\end{align}
we have
\begin{align}
\max_{j \in I} \| \bB_j^H\by \|_2^2 & \ge \rho_0 M[ 1-(d-1)\nu  ]^2s_u^2
-2\rho_0 M(N_a-1)[ 1+(d-1)\nu  ]d\mu_\bB s_l^2 \\
& \quad -2\sqrt {\rho_0 M}\tau [ (N_a-1)d\mu_\bB+[ 1+(d-1)\nu  ] ] s_l
\end{align}

Then we have
\begin{align} \label{condition1}
\max_{j \in I} \| \bB_j^H\by \|_2^2-\max_{j \notin I} \| \bB_j^H\by \|_2^2
 & \ge\rho_0 M[1-(d-1)\nu]^2s_u^2 -\tau^2-2\sqrt {\rho_0 M}N_ad\mu_\bB \tau s_u\\
&\quad -\rho_0 Md\mu_\bB\{ 2(N_a-1)[ 1+(d-1)\nu  ]s_l^2  +N_a^2d\mu_\bB s_u^2  \}\\
&\quad -2\sqrt {\rho_0 M} \tau  \{ (N_a-1)d\mu_\bB +[ 1+(d-1)\nu  ]  \}s_l
\end{align}

By \eqref{condition1}, \lemref{lem.bs} is proved.
\end{IEEEproof}

\begin{lemma}\label{lem.cir}
Suppose $\bu$ is a $MT$-dimensional circular symmetric Gaussian random vector
of zero mean and $\bI_{MT}$ covariance matrix, then
\begin{equation}
P\{ \ttau \ge \| \bB_j^H\bu
\|_2\} \ge 1-e^{-\varsigma^2}\sum_{k = 0}^{d-1} \frac{(
{{\varsigma^2}})^k} {k!}
\end{equation}
where $\varsigma = \ttau /\sqrt {1+({d-1})\nu }$.
\end{lemma}

\begin{IEEEproof}[Proof of \lemref{lem.cir}]
Suppose vector $\tbu = \sqrt 2 \bu$, then $\tbu$ satisfies the distribution of
$\mathcal{CN}( 0,2\bI)$. By \cite{re29}, $\| \tbu \|_2^2$ is a chi-squared
random variable with $2d$ degrees of freedom, then probability
\begin{equation}
Pr\{ \| \bu \|_2^2 \ge t^2 \}=Pr\{ \| \tbu \|_2^2 \ge 2t^2 \} = \frac{\Gamma (d,t^2)}{\Gamma (d)}
\end{equation}
where series expansion of $\Gamma (a,z)$ in \cite{re30} gives that
\begin{equation}
\Gamma (d,t^2)=\frac{e^{-t^2}}2t^2[ (\sqrt 2 t)^{2d}+(2d-2)(\sqrt 2 t)^{2d-2}+ \cdots +(2d-2)!!(\sqrt 2 t)^2 ]
\end{equation}

By using the double factorial notation: $n!! = \prod_{0 \le i \le \lfloor n/ 2
\rfloor } {({n-2i})}$, and $\Gamma (d) = (d-1)!$, we will obtain
\begin{align}  \label{regular}
\frac{\Gamma (d,t^2)}{\Gamma (d)} &=e^{-t^2}[ \frac{(t^2)^{d-1}}{(d-1)!}+\frac{(t^2)^{d-2}}{(d-2)!}+\frac{(t^2)^{d-3}}{(d-3)!}+ \cdots +\frac{(t^2)^0}{0!} ]\\
&=e^{-t^2}\sum_{k = 0}^{d-1} \frac{(t^2)^k}{k!}
\end{align}
which yields the lemma.

Consider the event $\ttau \ge \| \bB_j^H\bz \|_2$, as in \cite{re24},
$\bB_j^H\bz$ is a $d$-dimensional Gaussian random vector with zero mean and
$\bB_j^H\bB_j$ covariance matrix. Therefore, vector $\bu = (
\bB_j^H\bB_j)^{-1/2}{\bB_j^H}\bz$ is also a $d$-dimensional circular symmetric
Gaussian random vector of mean zero and covariance $\bI_d$. Then by
\lemref{lem.cir} and \reref{lem.coh} it is easy to demonstrate that
\begin{align}
\Pr \{ \| \bB_j^H\bz \|_2^2 \le \ttau^2 \}
&=\Pr \{ \| (\bB_j^H\bB_j)^{1/2}\bu \|_2^2 \le \ttau^2 \}\\
 &\ge \Pr \{ \| (\bB_j^H\bB_j) \|\cdot \| \bu \|_2^2 \le \ttau^2 \}\\
 &\ge \Pr \{ \| \bu \|_2^2 \le \frac{\ttau^2}{1+(d-1)\nu } \}\\
  &= 1-e^{-\varsigma^2}\sum_{k = 0}^{d-1} \frac{(\varsigma^2)^k}{k!}
\end{align}
where $\varsigma = \ttau /\sqrt {1+(d-1)\nu } $.
\end{IEEEproof}

With lemmas given above, we can prove \thref{th.1} next.

\begin{IEEEproof}[Proof of Theorem 1]
By the same induction in proof of \cite[Theorem 5]{re24}, when
\eqref{th_variables} is satisfied which verifies \eqref{okay_0}, for each
iteration of BOMP, an most likely active user will be selected. Therefore, the
actual support of the nonzero blocks can be correctly confirmed. Gathering all
these $K$ selected blocks to form a set, say $\hI$, we have $| \hI | = K$, and
$I \subseteq \hI$, then
\begin{align}
\| \hbs_{BOMP}-\bs \|_2^2
&=\| (\sqrt {\rho_0 M} \bB_{\hI})^\dag \by-\bs \|_2^2\\
&=\| (\sqrt {\rho_0 M} \bB_{\hI})^\dag \bz \|_2^2\\
% &\le {\| (\rho_0 M\bB_{\hI}^H\bB_{\hI})^{-1} \|^2}\| \sqrt {\rho_0 M} \bB_{\hI}^H\bz \|_2^2\\
 &\le (\rho_0 M)^{-1}\| (\bB_{\hI}^H\bB_{\hI})^{-1} \|^2\sum_{j \in \hI} \| \bB_j^H\bz \|_2^2 \\
 &\le \frac{K\tau^2}{[1-(d-1)\nu -(K-1)d\mu_\bB]^2\rho_0 M}
\end{align}
where $\tau = \max_{1 \le j \le N} \| \bB_j^H\bz \|_2$ and \eqref{lem.coh3} in
\reref{lem.coh} are used. The theorem is thus established.
\end{IEEEproof}

\ifCLASSOPTIONcaptionsoff
  \newpage
\fi

\end{document}